\documentclass[prl,twocolumn,showpacs,citeautoscript,superscriptaddress]{revtex4}

\usepackage{amsmath}
\usepackage{graphicx}

\begin{document}

\preprint{\today}

\title{\boldmath Pressure-driven orbital reorientation and change in Mott-Hubbard gap in YTiO$_{3}$}

\author{I. Loa}
\email[Corresponding author:~E-mail~]{I.Loa@fkf.mpg.de}
\author{X. Wang}
\author{K. Syassen}
\affiliation{Max-Planck-Institut f{\"u}r Festk{\"o}rperforschung,
Heisenbergstr.\ 1, D-70569 Stuttgart, Germany}

\author{H. Roth}
\author{T. Lorenz}
\affiliation{Universit{\"a}t zu K{\"o}ln, {II}. Physikalisches Institut,
Z{\"u}lpicher Str.\ 77, D-50937 K{\"o}ln, Germany}

\author{M. Hanf\/land}
\affiliation{European Synchrotron Radiation Facility,
        BP 220, F-38043 Grenoble, France}

\author{Y.-L. Mathis}
\affiliation{ANKA/ISS, Forschungszentrum
Karlsruhe, PF 3640, D-76021 Karlsruhe, Germany}

\date{\today}

\begin{abstract}
We investigate the crystal structure of YTiO$_{3}$\ at high pressures up to 30~GPa
by synchrotron x-ray powder diffraction ($T=295$~K). The variation of the
\mbox{Ti--O} bond lengths with pressure evidences a distinct change in the
distortion of the TiO$_{6}$ octahedra at around 10~GPa, indicating a
pressure-driven spatial reorientation of the occupied Ti~$3d(t_{2g})$
orbitals. The pressure-induced reduction of the optical  band gap of YTiO$_{3}$\ is
determined quantitatively by mid-infrared synchrotron micro-spectroscopy  and
discussed in terms of bond length and orbital orientation changes.
\end{abstract}

\bigskip

\pacs{%
%
 61.50.Ks,  
 61.10.Nz, 
 71.30.+h 
}

\maketitle

\begin{figure}[tb]
     \centering
     \includegraphics[width=0.85\hsize]{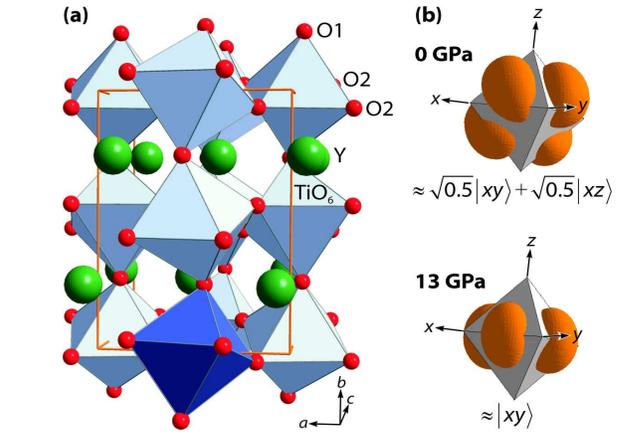}
     \caption{(a) Perovskite-type (GdFeO$_{3}$)
     crystal structure of YTiO$_{3}$\ (space
     group $Pnma$) \cite{MNG79}. (b) Orientation of the titanium $3d^1(t_{2g})$
     wavefunction at 0 and 13 GPa as derived in this work from the TiO$_{6}$
     octahedral distortion. The orbitals shown refer to the Ti site
     highlighted in (a). The octahedra in (b) illustrate the octahedral
     environment of the orbitals, but they do not match the size of the actual
     TiO$_{6}$ octahedra.}
     \label{fig:YTO_structure}
\end{figure}

Orbital ordering, fluctuation and excitation phenomena of $3d$ electrons in
transition metal perovskites have attracted much interest. For $e_g$ electron
systems like LaMnO$_{3}$\ [electron configuration $3d^4(t_{2g}^3 e_g^1)$], lifting of
energetic degeneracies due to lattice distortions (Jahn-Teller effect) and
concomitant spatial ordering of the occupied $e_g$ orbitals are well known. In
contrast, the $t_{2g}$ states in systems like rare-earth titanates
[configuration $3d^1(t_{2g}^1)$] have often been assumed to be degenerate and
thus in disfavor of orbital ordering. YTiO$_{3}$\ [Fig.~\ref{fig:YTO_structure}(a)]
is a prototypical Mott-Hubbard insulator with ferromagnetic ground state,
where the question of $t_{2g}$ orbital ordering has been studied in some
detail. Static ordering was predicted theoretically \cite{ST98,MF96} and
confirmed in nuclear magnetic resonance \cite{ITTM99} and several other
experiments \cite{AIEM01,NWKM02,ITSH04}. The mechanism and energetics of the
ordering were studied in theoretical work \cite{MKS99,PBPL04,MI04a,PYNA05p}.
In contrast to these experimental and theoretical findings, the experimentally
observed spin-wave excitation spectrum was taken as evidence for strong
orbital fluctuations \cite{UKOR02,KO02}. However, an alternative explanation,
compatible with orbital order, has been proposed \cite{PYNA05p}.

Altogether, these recent studies raise a number of questions. How robust is
orbital ordering in $t_{2g}$ electron systems like YTiO$_{3}$? Which factors
determine the orbital ordering in a particular compound? Can it be tuned by
external parameters such as high pressure? How does orbital order relate to
other physical properties, e.g., the electronic and magnetic excitation
spectra? The orbital polarizations in the rare-earth titanates manifest
themselves in small distortions of the TiO$_{6}$ octahedra \cite{MI04a}. It is
therefore possible to study the orbital ordering of YTiO$_{3}$ by structural
methods.

In this letter, we demonstrate that the TiO$_{6}$ octahedra remain distorted
under pressure, showing the persistence of orbital ordering. This order is,
however, not an immutable property: experimental evidence is presented that
the Ti $3d (t_{2g})$ orbitals undergo a \emph{pressure-driven reorientation}.
To relate the structural/orbital changes to the electronic properties of YTiO$_{3}$,
we determine quantitatively (by mid-infrared spectroscopy) how the
Mott-Hubbard gap decreases under pressure.


\begin{figure}[t]
     \centering
     \includegraphics[width=0.62\hsize]{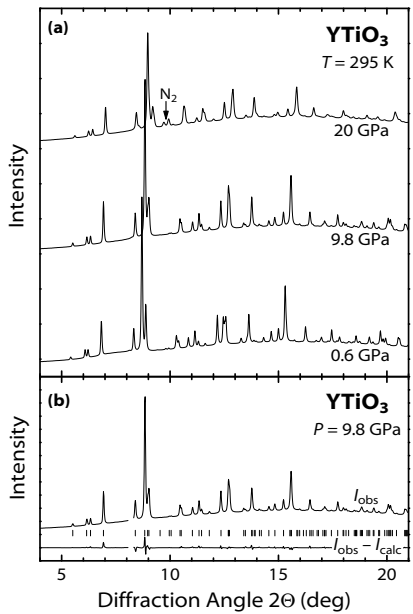}
     \caption{\small(a) Selected x-ray diffraction diagrams of YTiO$_{3}$\ for pressures
     up to 20~GPa ($T=295$~K, $\lambda = 0.41$~{\AA}). (b) Rietveld refinement of
     data recorded at 9.8~GPa. $I_{\rm obs}$ and $I_{\rm calc}$ denote
     observed and calculated diffraction intensities, respectively. Markers
     show the calculated peak positions.
     }
     \label{fig:YTO_diffpat}
\end{figure}

Angle-dispersive x-ray powder diffraction experiments were performed at the
beamline ID09A of the European Synchrotron Radiation Facility in Grenoble. A
fine powder was produced from an  YTiO$_{3}$\ crystal (Curie temperature $T_C =
28$~K) grown by a floating zone technique \cite{CLBM03}. The sample was
pressurized in a diamond-anvil cell using condensed nitrogen as the
pressure-transmitting medium. Two-dimensional diffraction images were recorded
with an image plate detector and converted to intensity-vs-2$\theta$
dif\-fracto\-grams [Fig.~\ref{fig:YTO_diffpat}] by numerical integration
\cite{soft:fit2d}. The structural parameters were determined by means of
Rietveld refinements \cite{soft:GSAS,soft:EXPGUI}. In contrast to the common
practice, we calculated the statistical uncertainties of the diffraction
intensities from the intensity variations along the individual diffraction
rings, rather than assuming basic counting statistics ($\Delta I
\propto \sqrt{I}$). Using the experimentally determined uncertainties as weights in
the Rietveld fitting process turned out to be essential for an accurate
determination of the oxygen atomic positions. Mid-infrared transmission
experiments were conducted at the infrared beamline of the synchrotron ANKA in
Karlsruhe. The spectra were recorded on 40~\mbox{\ensuremath{\mu}m}\ thick YTiO$_{3}$\ crystals with a
Bruker IFS66v/S Fourier transform spectrometer equipped with an MCT detector
and a microscope using mirror objectives. Synthetic type-IIa diamond anvils
were used in the infrared experiments. Condensed nitrogen and solid KCl were
employed as pressure media. Pressures were determined in all experiments with
the ruby luminescence method \cite{ruby:method}.

\begin{figure}
     \includegraphics[width=0.95\hsize]{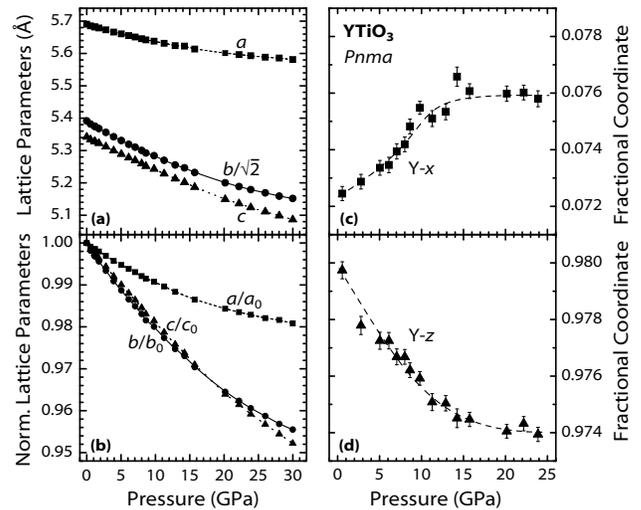}
     \caption{Structural parameters of YTiO$_{3}$\ as a function of pressure. (a) Lattice parameters,
     (b) lattice parameters normalized to their respective zero-pressure
     values, (c, d) $x$ and $z$ fractional coordinates of Y. The lattice
     parameter $b$ was scaled by $\sqrt{2}$ in (a) to obtain a pseudo-cubic
     representation.}
     \label{fig:YTO_StructureInfo}
\end{figure}

The diffraction diagrams of YTiO$_{3}$\ (Fig.~\ref{fig:YTO_diffpat}) and the
measured lattice parameters [Fig.~\ref{fig:YTO_StructureInfo}(a,\,b)] change
continuously with pressure. The ambient-pressure perovskite-type crystal
structure of YTiO$_{3}$\ is thus stable under compression up to at least 30~GPa. The
compressibility exhibits a distinct anisotropy with the $a$ direction being
only about half as compressible as the $b$ and $c$ directions
[Fig.~\ref{fig:YTO_StructureInfo}(b)]; the orthorhombic strain increases with
pressure. A Birch equation of state was fitted to the pressure-volume data up
to 30~GPa to determine the bulk modulus $B_0 = 163(6)$~GPa and its pressure
derivative at zero pressure $B' = 8.5(10)$; the zero-pressure volume was fixed
at the measured value of $V_0 = 231.78(2)$~{\AA}$^{3}$. The increasing distortion
is evident also from the variation of the yttrium coordinates $x$ and $z$ with
pressure shown in Fig.~\ref{fig:YTO_StructureInfo}(c,\,d) [$y=1/4$ by
symmetry]. Under pressure, the yttrium ions clearly move further away from the
``ideal'' position $(0, 1/4, 1)$ in an undistorted (cubic) perovskite. This
shift levels off, however, at around 15~GPa.

\begin{figure}[t]
     \centering
     \includegraphics[width=0.6\hsize]{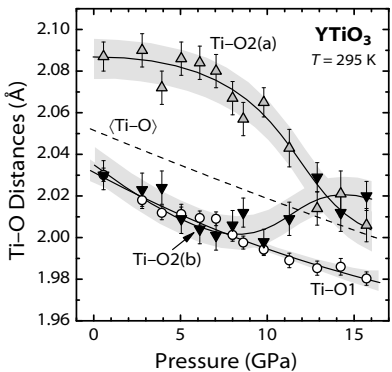}
     \caption{\small Ti--O bond lengths in YTiO$_{3}$\ as a function of
     pressure (symbols). $\langle\mbox{Ti--O}\rangle$ denotes the average bond
     length (dashed line). Grey bands indicate estimated confidence bands;
     solid lines are guides to the eye.}
     \label{fig:YTO_TiO-dist}
\end{figure}

The variations of the Ti--O distances with pressure
(Fig.~\ref{fig:YTO_TiO-dist}) represent the most important structural
information. At ambient pressure, the long Ti--O2(a) distance exceeds the two
shorter ones by $\sim$3\% [Ti--O2(a) and Ti--O2(b) denote the two distinct
Ti--O2 distances]. This type of distortion persists up to about 8~GPa. When
increasing the pressure further, the initially short Ti--O2(b) distance
\emph{lengthens} with increasing pressure, while Ti--O2(a) shortens markedly.
As a result, the two Ti--O2 bond lengths become (i) nearly equal at 13~GPa and
(ii) distinctly larger than the Ti--O1 distance.

As has been described in detail before \cite{AIEM01,MI04a}, the deformation of
the TiO$_{6}$ octahedra reflects the spatial orientation of the $t_{2g}$
wavefunction. To facilitate the discussion, we introduce a \emph{local}
coordinate system at each Ti site with the $x$, $y$, and $z$ axes parallel to
the \mbox{Ti--O2(a)},
\mbox{Ti--O2(b)}, and \mbox{Ti--O1} bonds, respectively
[Fig.~\ref{fig:YTO_structure}]. The relative Ti--O distances at 0~GPa (long
along $x$; short along $y$ and $z$) indicate nearly equal occupancy of the
\ensuremath{|xy\rangle} and \ensuremath{|xz\rangle} orbitals. Thus, the
eigenfunction is $\approx \sqrt{0.5}\ensuremath{|xy\rangle} +
\sqrt{0.5}\ensuremath{|xz\rangle}$, in agreement with the experimental results
\cite{ITTM99,AIEM01,NWKM02,ITSH04} and theoretical studies
\cite{MF96,ST98,MKS99,PBPL04,MI04a}.

At 13~GPa, the TiO$_{6}$ octahedra are characterized by two equally long
Ti--O2 bonds along $x$, $y$ and a shorter Ti--O1 distance along $z$, i.\ e.,
the octahedra are compressed along $z$. Such a distortion indicates
predominant occupation of the \ensuremath{|xy\rangle} orbital. In fact, the relative
\mbox{Ti--O} bond lengths in YTiO$_{3}$\ above 13~GPa are very similar to those in
SmTiO$_{3}$ at ambient pressure, where there is theoretical support for a
predominant occupation of the \ensuremath{|xy\rangle} orbitals \cite{MI04a}. The changes in
the Ti--O bond lengths observed in YTiO$_{3}$\ thus provide evidence of \emph{a
pressure-induced reorientation of the $t_{2g}$ orbitals} from the initial
``tilted state'' with the approximate wavefunction
$\sqrt{0.5}\ensuremath{|xy\rangle} + \sqrt{0.5}\ensuremath{|xz\rangle}$ to a
situation where essentially only the \ensuremath{|xy\rangle} orbital is
occupied, as shown in Fig.~\ref{fig:YTO_structure}(b).

\begin{figure}[t]
     \centering
     \includegraphics[width=0.85\hsize]{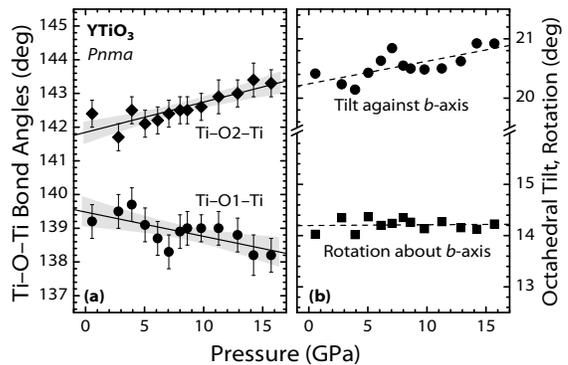}
     \caption{(a) Ti--O--Ti bond angles and (b) octahedral tilt and rotation
     angles in YTiO$_{3}$\ as a function of pressure. Lines are guides to the eye;
     the shades bands indicate confidence bands.}
     \label{fig:YTO_Angles}
\end{figure}

It is interesting to note that the orbital reorientation does not correlate
with anomalies in the \emph{average} Ti--O distance
[Fig.~\ref{fig:YTO_TiO-dist}], the lattice parameters
[Fig.~\ref{fig:YTO_StructureInfo}(a,\,b)], or the Ti--O--Ti bond angles
[Fig.~\ref{fig:YTO_Angles}]. In contrast, the change in the pressure-induced
shift of the Y ions at 10--15~GPa [Fig.~\ref{fig:YTO_StructureInfo}(c,\,d)]
coincides with the Ti $t_{2g}$ orbital reorientation.

For a qualitative discussion of the origin of the orbital reorientation under
pressure, we consider the results of recent theoretical investigations
\cite{MI04a,PYNA05p}. Essentially, one has to take into account two effects
that induce a splitting of the $t_{2g}$ states. Firstly, in the presence of
the GdFeO$_{3}$-type distortion, the crystal field of the Y ions lifts the
degeneracy of the Ti $t_{2g}$ states. This splitting will increase if the
Y--Ti distances are shortened, i.e., under pressure. Secondly, a Jahn-Teller
distortion due to the Ti-O interaction can modify or even dominate the
energetics of the Ti $t_{2g}$ states. Due to the stiffening of the lattice,
Jahn-Teller distortions will usually become less favorable under compression
\cite{LAGS01}. Application of high pressure is thus expected to tune the
balance between these two contributions. In terms of relative Ti--O bond
lengths, the situation in YTiO$_{3}$\ at 15~GPa is quite similar to that in
SmTiO$_{3}$ at ambient pressure. In the latter case, there exists no
Jahn-Teller distortion of the type encountered in YTiO$_{3}$, and the competition
between the crystal field of the rare-earth ions and that of the O ions was
reported to be rather balanced \cite{MI04b}. Quite likely, this scenario
applies also to YTiO$_{3}$\ at $\sim$15~GPa.

The similarity between YTiO$_{3}$\ and SmTiO$_{3}$ has important implications
regarding magnetism. In contrast to YTiO$_{3}$, SmTiO$_{3}$ is an antiferromagnet
\cite{KTT97}. In view of the orbital reorientation and the coupling between
spins and orbitals, one may speculate that YTiO$_{3}$\ becomes antiferromagnetic
under pressure. YTiO$_{3}$\ would then lose its unusual property of being
simultaneously ferromagnetic \emph{and} insulating. More generally,
application of pressure to YTiO$_{3}$\ may cause similar changes in the electronic
and magnetic properties as a substitution of Y by larger rare-earth ions.

\begin{figure}[t]
     \centering
     \includegraphics[width=0.95\hsize,clip]{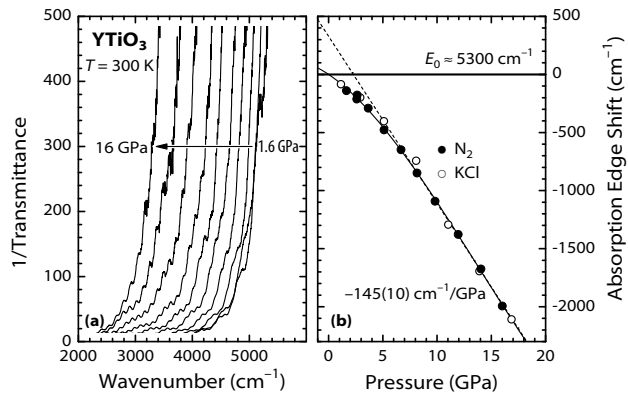}
     \caption{Mid-infrared optical absorption in YTiO$_{3}$\ under pressure
     ($T=295$~K).
     (a) 1/Transmittance spectra recorded at pressures of 1.6, 2.6,  3.7,
     5.1,  6.7,  8.2,  9.8,  12.0, 14.1, and 16.0~GPa (with nitrogen pressure
     medium). The modulations of the spectra originate from interferences.
     (b) Pressure-induced shift of the optical absorption edge determined at
     1/Transmittance = 300 from two experimental runs with condensed nitrogen and KCl
     pressure medium, respectively.}
     \label{fig:YTO_absorption}
\end{figure}

We turn now to the question of the electronic excitation spectrum of
YTiO$_{3}$\ under pressure and whether it is affected by the orbital
reorientation. Figure~\ref{fig:YTO_absorption} depicts spectra of the
reciprocal transmittance $1/T$ [which is related to the absorbance
$A=\log_{10}(1/T)$] for several pressures up to 16~GPa. In this
representation, the optical absorption edge and its pressure-induced red shift
are easily recognized.  To obtain quantitative results on the red shift, we
determine the energies where $1/T = 300$ (corresponding to an optical
conductivity of $\sim$10~$\Omega^{-1}\,\text{cm}^{-1}$). The ambient-pressure
optical gap thus determined at the \emph{onset of absorption} amounts to
$\sim$5300~cm$^{-1}$\ (0.7~eV), a number that is naturally somewhat smaller than the
previously reported values of 6500--8000~cm$^{-1}$\ (0.8--1.0~eV) deduced from
reflectance measurements \cite{ATT93,OKOA95}. The important information here
is the shift of the absorption edge [Fig.~\ref{fig:YTO_absorption}(b)] rather
than its absolute value. In the higher-pressure range, the optical band gap
decreases essentially linearly with pressure at a rate of $-145(10)$~cm$^{-1}$/GPa.
On the basis of a linear extrapolation of these data, the optical gap is
expected to close at a pressure in the order of 40~GPa.

The mid-IR optical absorption of rare-earth titanates has been attributed to
an excitation across the Mott-Hubbard gap \cite{ATT93,CTGG92}. The Coulomb
repulsion $U$ is not expected to change under pressure so that the reduction
in bandgap is a measure of the increase in bandwidth. The increasing
electronic bandwidth under pressure is usually interpreted in terms of the
Ti--O bond lengths and the Ti--O--Ti bond angles. In the present case, the
changes in the bond angles, i.e.\ $\pm 1$\mbox{$^\circ$}\ up to 16~GPa, are relatively
small compared to related perovskites like rare-earth nickelates \cite{Amb03}
or LaMnO$_{3}$ \cite{LAGS01}. In fact, the \emph{average} bond angle is
essentially pressure-insensitive due to the opposite changes of Ti--O1--Ti and
Ti--O2--Ti. Therefore, we suppose that the variation in the average Ti--O bond
length $d$ dominates the reduction of the Mott-Hubbard gap $E_g$ under
pressure. This can be quantified in terms of a gap deformation potential, $\ensuremath{\mathrm{d}}
E_g / \ensuremath{\mathrm{d}} \ln d = 11(2)$~eV in the region of linear change.

Figure~\ref{fig:YTO_absorption}(b) evidences an unexpected nonlinear change of
the optical gap with pressure: the initial slope of the absorption edge shift
is only about half as large as compared to the region above 10~GPa. The
increase in slope up to $\sim$10~GPa can hardly be correlated with changes in
the average bond length or the bond angles. This seems to indicate that the
pressure-induced band gap closure depends not only on the bond lengths and
angles, as is usually assumed, but also on the orbital state of the transition
metal ion. Such view finds support in recent theoretical work \cite{PYNA05p},
where under \emph{isotropic compression} (that is, without orbital
reorientation) the electronic gap closure is expected to occur only at
$\sim$100~GPa, i.e., at about twice the experimental estimate.

In summary, the pressure-induced structural changes in YTiO$_{3}$\ were studied in
detail by synchrotron x-ray diffraction. Distinct changes in the distortion of
the TiO$_{6}$ octahedra around 10~GPa are attributed to a spatial
reorientation of the occupied Ti~$3d(t_{2g})$ orbitals. There is a structural
similarity between
YTiO$_{3}$\ at 15~GPa and SmTiO$_{3}$ at ambient pressure, which suggests that YTiO$_{3}$\
may lose its unusual property of being both ferromagnetic \emph{and}
insulating by turning antiferromagnetic under compression. The combination of
infrared transmission and x-ray diffraction experiments made it possible to
obtain direct, quantitative information on the change of the electronic
bandgap/bandwidth in a rare-earth transition metal perovskite together with a
detailed knowledge of the associated structural changes. These results provide
a framework for testing theoretical methods and models of the electronic
structure of the $t_{2g}$-band titanates in the vicinity of the
insulator-metal borderline.

\acknowledgments
We acknowledge stimulating discussion with O.~K.~Andersen, K. Held, and A.
Yamasaki. We thank F.~X. Zhang for support in the early stage of this study
and M.~S{\"u}pfle for technical support at ANKA.

\bibliographystyle{prsty}



\end{document}